\documentclass[pdftex,pre,floats,twoside,twocolumn,amsmath,amssymb,floatfix]{revtex4}
\usepackage{graphicx}
\usepackage{natbib}
\usepackage{color}
\definecolor{darkgray}{rgb}{0.25,0.25,0.25}
\definecolor{darkred}{rgb}{0.89,0.10,0.11}
\definecolor{darkblue}{rgb}{0.12,0.39,0.62}
\usepackage{url}
\usepackage[pdftex,breaklinks=true,colorlinks=true,citecolor=black,linkcolor=black,menucolor=black,urlcolor=darkblue,pdfborder={1 0 0}]{hyperref}
\hypersetup{pdftitle={Ranking and clustering of nodes in networks with smart teleportation},pdfauthor={Renaud Lambiotte and Martin Rosvall, 2012}}
\begin{document}
\makeatletter
\renewcommand\@biblabel[1]{#1.}
\makeatother
	
\title{Ranking and clustering of nodes in networks with smart teleportation}	
	
\date{\today}

\author{R. Lambiotte}
\email{renaud.lambiotte@fundp.ac.be}
\homepage{http://www.lambiotte.be/}
\affiliation{Department of Mathematics and Naxys, University of Namur, 5000 Namur, Belgium}

\author{M. Rosvall}
\email{martin.rosvall@physics.umu.se}
\homepage{http://www.tp.umu.se/~rosvall/}
\affiliation{Integrated Science Lab, Department of Physics, Ume{\aa} University, SE-901 87 Ume{\aa}, Sweden}

\begin{abstract}
Random teleportation is a necessary evil for ranking and clustering directed networks based on random walks. Teleportation enables ergodic solutions, but the solutions must necessarily depend on the exact implementation and parametrization of the teleportation. For example, in the commonly used PageRank algorithm, the teleportation rate must trade off a heavily biased solution with a uniform solution. Here we show that teleportation to links rather than nodes enables a much smoother trade-off and effectively more robust results. We also show that, by not recording the teleportation steps of the random walker, we can further reduce the effect of teleportation with dramatic effects on clustering.
\end{abstract}

\maketitle

\section*{\large Introduction}

Random walks play a preponderant role in network theory \cite{boccaletti2006complex} and are at the heart of popular metrics measuring the effect of network topology on patterns of flows through the nodes. Defined as the expected density of random walkers on a node at stationarity, PageRank provides a non-local measure of centrality and is perhaps the most important and influential application of random walks \cite{brin1998anatomy}.
First introduced to rank pages on the Web, PageRank \cite{brin1998anatomy,langville2006google}, or variations of it \cite{haveliwala2003topic,Goncalves09wsdm,delvenne2011centrality}, has now been adopted to rank the importance of nodes in a broad range of systems, e.g., in citation networks \cite{bergstrom2008eigenfactor,chen2007finding}, food-webs \cite{allesina2009googling}, and sports \cite{radicchi2011best}. Similarly, in the field of community detection, more and more methods are based on the notion that networks often describe systems characterized by flow and the intuitive idea that random walkers should be trapped for long times in good communities. This idea led to the design of quality functions for network partitioning such as the so-called map equation \cite{rosvall2008maps} or stability \cite{delvenne2010stability}, which naturally take into account the constraints imposed by network topology on dynamical processes.

Random walk-based methods are appealing because of their nice mathematical properties, their ability to explore the system at multiple scales, and their intuitive interpretation of how real flows of people, money, information, etc.\ take place in empirical networks \cite{borgatti2005centrality,lambiotte2011flow}. However, most methods suffer from an important drawback: they are defined only at stationarity, a state that is either trivial, non-uniquely defined, or never reached in a majority of empirical systems. To circumvent this problem, mathematical tricks have been proposed to make the dynamics ergodic, even when the underlying network is not strongly connected. The most prominent procedure allows walkers to randomly teleport across the system, and thus to occasionally free themselves from the actual topology. Unfortunately, teleportation brings its own share of problems. For example, with teleportation, the ranking of nodes or their clustering into communities depend not only on the topological properties of the system, but also on the exact implementation of the artificial teleportation process.

\begin{figure}[tpb]
    \centering
   \includegraphics[width=\columnwidth]{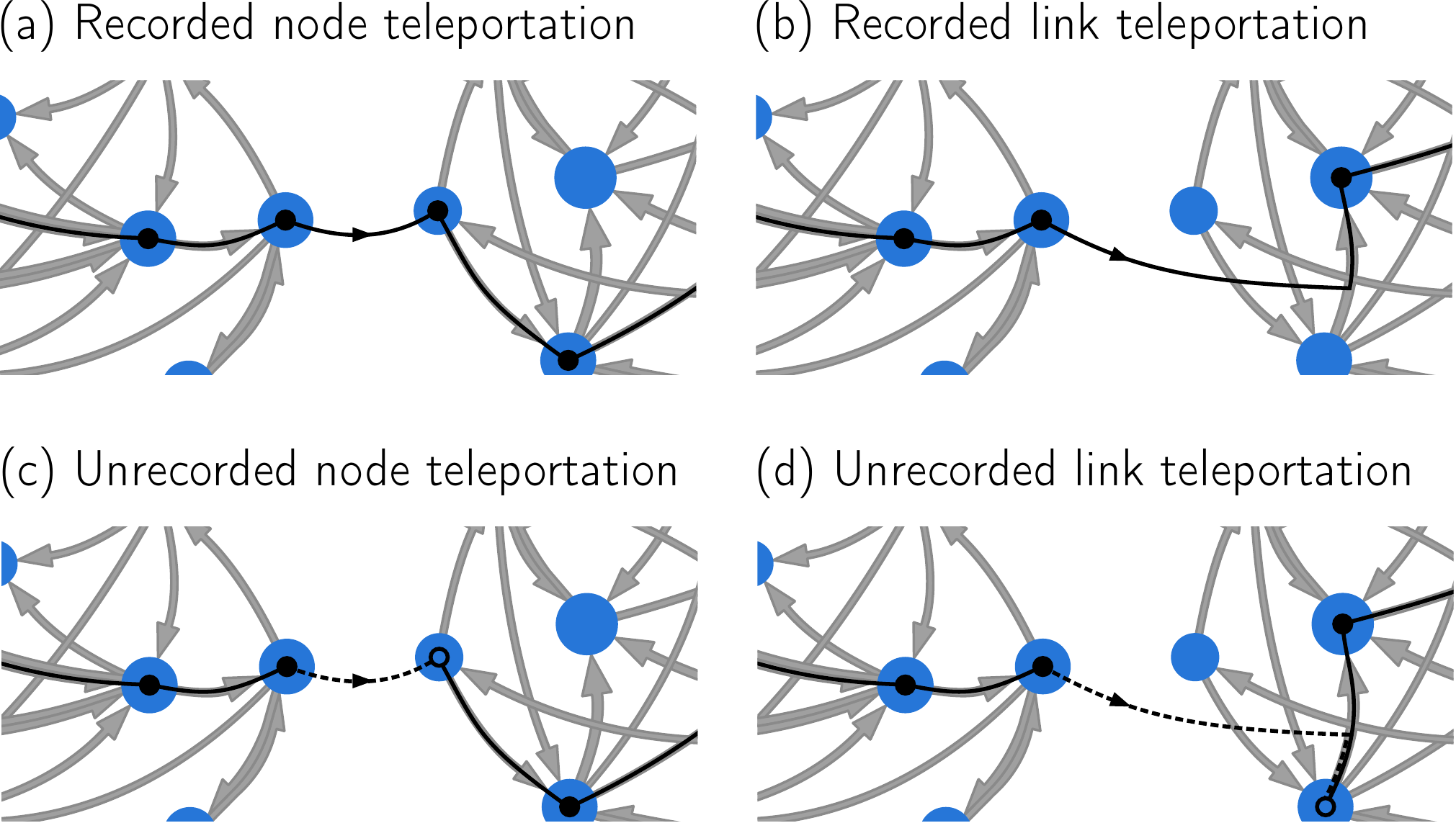}
    \caption{.pdfCommon and smart teleportation in networks. (a) Recorded node teleportation is the commonly used teleportation scheme. Both steps along links and teleportation steps contribute to node visit rates for ranking and transition rates for clustering, and nodes are the targets of teleportation. (b) In recorded link teleportation, all steps contribute and links are the targets of teleportation. (c) In unrecorded node teleportation, only steps along links (solid lines and filled circles) contribute, and not those due to teleportation (dashed line and open circle). (d) In unrecorded link teleportation, only steps along links contribute and links are the targets of teleportation. \label{schematicwalk}}
\end{figure}

The goal of this paper is to propose and evaluate different ways to minimize the effect of teleportation on random walk-based metrics and methods. To do so, we explore two different but related possibilities for smart teleportation. In order to make rankings more robust, our first approach modifies the targets of teleportation steps, the so-called preference vector. In order to make clusterings more robust, our second approach modifies which steps contribute to transition rates between nodes and only counts steps along links and not teleportation steps.

Figure \ref{schematicwalk} describes the different teleportation schemes. In general, the probability of landing on a node after a teleportation depends on some of the topological properties of the network. Standard teleportation, which we call \emph{recorded node teleportation}, is recovered when the preference vector is uniform, i.e., the probability to land on each node is the same (see Fig.~\ref{schematicwalk}(a)). In this paper, for ranking we argue for the use of \emph{recorded link teleportation}, where the preference vector is proportional to the in-strength of the nodes (see Fig.~\ref{schematicwalk}(b)), and is equivalent to teleporting to links instead of nodes. For clustering we argue for the use of teleportation without recording. No teleportation steps are recorded when walkers teleport uniformly to nodes in \emph{unrecorded node teleportation} (see Fig.~\ref{schematicwalk}(c)) and to nodes proportionally to their out-strength in \emph{unrecorded link teleportation} Fig.~\ref{schematicwalk}(d)).

The difference between the \emph{recorded} and \emph{unrecorded} schemes stems from the fact that only steps along links contribute to transition rates between nodes, which we will show to be crucial for improving community detection. Below, we study the mathematical relations and differences between the four teleportation schemes illustrated in Fig.~\ref{schematicwalk}. We show that the incorporation of appropriate topological elements into teleportation leads to desirable properties in different limit scenarios, and provides an interesting connection between local and non-local centrality measures. Numerical simulations also confirm that the effect of teleportation on ranking and clustering can be significantly reduced, with important applications for mining the large-scale organization of complex networks.

\section*{\large Mathematics of teleportation}

We focus on weighted and directed networks described by the $N \times N$ adjacency matrix $W_{ij}$, where $N$ is the number of nodes in the system and $W_{ij}$ is the weight of the link from $j$ to $i$. The total in- and out-strengths of node $i$ are defined as $w_{i}^{\rm in}=\sum_{j} W_{ij}$ and $w_{i}^{\rm out}=\sum_{j} W_{ji}$, respectively. The total weight of all links $W$ is given by $W=\sum_i w_{i}^{\rm out} = \sum_i w_{i}^{\rm in}$.
In the case of unweighted networks, the adjacency matrix is equal to 1 if there is a link going from $j$ to $i$ and 0 otherwise. Moreover, $w_{i}^{\rm in}$ and $w_{i}^{\rm in}$ correspond to the in- and out-degrees of node $i$.

\subsection*{Standard teleportation}

The dynamical properties of an unbiased random walker on a network are determined by the spectral properties of the transition matrix $T_{ij}=W_{ij}/w_{j}^{\rm out}$, which drives the time-evolution of the expected density $p_i$ of walkers at node $i$
 \begin{align}
\label{discretedirected}
p_{i;t+1} = \sum_{j} T_{ij} p_{j;t}.
\end{align}
The steady-state density of walkers is given by the dominant eigenvector of $T_{ij}$, denoted by $\pi_i$, which defines the PageRank of node $i$. Asymptotic convergence towards this solution and its uniqueness are ensured only if the network is strongly connected and aperiodic, a situation that rarely occurs in empirical networks. In order to regularize this situation, several tricks have been proposed in the literature, the most common being to allow for teleportations through the network. In its simplest instance, walkers either follow links with probability $\alpha$ or teleport to a random location with probability $1-\alpha$. 

Random walks with teleportation are driven by the rate equation
\begin{align}
\label{preference}
p_{i;t+1} = \alpha \sum_{j} T_{ij} p_{j;t} + (1-\alpha) v_i,
\end{align}
where the preference vector $v_i$, subject to the constraint $\sum v_i=1$, determines the frequency at which walkers teleport to node $i$. In general, the random process (\ref{preference}) converges towards a unique steady-state solution for any $\alpha < 1$. Moreover, the stationary solution of (\ref{preference}) is a function of $v_i$ and of the teleportation probability $1-\alpha$, formally given by 
 \begin{align}
\label{formal}
\pi_{i;\alpha} = (1-\alpha) \sum_j (I-\alpha T)^{-1}_{ij} v_j
\end{align}
where the dependence on $\alpha$ has been made explicit.
This solution can be Taylor expanded in terms of $\alpha$ to provide the expression \cite{boldi2005pagerank,brinkmeier2006pagerank}:
 \begin{align}
\label{formalB}
\pi_{i;\alpha} = v_i + \sum_{k=1}^{\infty} \alpha^k  \sum_j \left( T^k_{ij} - T^{k-1}_{ij}  \right) v_j,
\end{align}
an expression that clearly shows the non-local nature of PageRank, as it is made of terms associated with paths of any length $k$ in the network.

In the above expressions, we have implicitly assumed that each node has at least one outgoing link, such that $w_{i}^{\rm out} > 0$, $\forall i$, and that the transition matrix $T_{ij}$ preserves probability, i.e. $\sum_i T_{ij}=1$. In systems where this condition is not fulfilled, it is usual to impose a teleportation step every time a walker arrives on a dangling node $j$ without out-links. Mathematically, this corresponds to replacing the $j^{\rm th}$ column of $T_{ij}$, only made of zeros, by the preference vector $v_i$. 
For the sake of simplicity, but without loss of generality, in the following mathematical analysis we will assume that the system does not contain dangling nodes.

\subsubsection*{Limitations of standard teleportation}

Random walks with teleportation have the advantage of making the dynamics ergodic and thus ensure the existence of a well-defined, asymptotic, steady-state solution. However, due to its artificial nature and the extra parameter, the teleportation process also raises a series of fundamental questions  \cite{boldi2005pagerank}. While a random walk is a good proxy for diffusion in a broad range of networked systems, teleportation can only be viewed as a mathematical trick in the absence of real-world interpretation. Moreover, even when such an interpretation is plausible, e.g., for individuals browsing the Web and occasionally jumping to a new page without following a hyper-link, selecting a proper value of $\alpha$ and an expression for $v_i$ is problematic. 

Most research tends to overlook these issues and use the standard value $\alpha=0.85$ and the uniform preference vector $v_i=1/N$, i.e., a walker randomly teleports on any node, independently of any intrinsic or topological properties. This choice of preference vector leads to the recorded node teleportation illustrated in Fig.~\ref{schematicwalk}(a).
Yet it has been shown that the stationary solution $\pi_i$ can radically change when $\alpha$ is modified \cite{pretto2002theoretical,langville2004deeper,avrachenkov2008singular}. 
This dependence is clear when rewriting the formal solution (\ref{formalB}) with $v_i=1/N$
 \begin{align}
\label{formalK2}
\pi_{i;\alpha} = \frac{1}{N} + \sum_{k=1}^{\infty} \frac{\alpha^k}{N}  \sum_l  T^{k-1}_{il} \sum_{j} T_{lj} \left( \frac{w_l^{\rm in}- w_j^{\rm out} }{w_l^{\rm in} } \right).
\end{align}
The leading contribution for small $\alpha$ makes PageRank uniform, thereby diluting the structural differences between nodes, whereas differentiation emerges when $\alpha$ is increased.
The contribution of each path of length $k$ is proportional to the difference between in- and out-strengths around links, i.e., $w_l^{\rm in}- w_j^{\rm out}$. If the network is strongly connected, it is instructive to note that all of these contributions vanish only when the network is regular, i.e., $w_i^{\rm in} = w_i^{\rm out} = W/N$, $\forall i$. This version of PageRank is thus expected to depend on $\alpha$ except in this trivial case. For these reasons, it is important to improve our understanding of the sensitivity due to teleportation and to identify adequate values of $\alpha$. So far, the rule of thumb has been to choose values close to $1$, in order to minimize the effect of teleportation on the random walk process, but not too close, because calculations become prohibitively expensive and unstable in this limit. 
Similarly, the importance of the preference vector $v_i$ is ignored in a majority of studies, despite the fact that, in general, no individual choice is better than any other one, and that different choices seem more realistic in different types of systems \cite{baeza2006generalizing,allesina2009googling}. For instance, in systems where the size of the nodes is heterogeneous, e.g., scientific journals publish different numbers of articles, the preference vector can be chosen proportional to the size of the nodes \cite{bergstrom2008eigenfactor}.

\subsection*{Smart Teleportation}

Teleportation can be seen as a mean-field process where walkers jump towards any node $i$ with a probability $v_i$, independently of the underlying network topology. 
Our aim is to reduce the noise induced by teleportation in order to produce a more faithful description of the system, and to minimize its dependence on the value of $\alpha$.

\subsubsection*{Recorded link teleportation}

Our first approach takes advantage of the ability to choose an appropriate preference vector to improve the stability of $\pi_{i;\alpha}$.
In the PageRank literature, the preference vector $v_i$ has been introduced as a way to incorporate non-structural properties into the algorithm and to fine-tune PageRank to the particular taste or interest of a user. Here we propose instead to select a preference vector based on topological properties of nodes, with the aim of minimizing the effect of teleportation on dynamics. In the ideal case of a strongly connected and aperiodic network, an appealing solution is to take $v_i$ proportional to $\pi_i$, solution of Eq.~\ref{discretedirected}.  In that case, $\pi_i$ is well defined and it is easy to show that $\pi_{i;\alpha}=\pi_i$ for all values of $\alpha$. The question of picking a particular value of $\alpha$ thus becomes unimportant.  
 
The previous example is trivial because teleportation is not necessary to make the original dynamics ergodic. Nonetheless, it provides useful hints on how to address the general problem: one should aim for a preference vector likely to be close to $\pi_i$. 
To do so, we propose the use of
 \begin{align}
 \label{pref}
v_i = \frac{w_i^{\rm in}}{W},
\end{align}
inspired by the observations that in-strength is statistically correlated to PageRank in random networks and that both quantities are equivalent, up to an additive constant, in the mean-field approximation \cite{fortunato2008approximating,litvak2007degree}. This process is equivalent to selecting a link at random  proportionally to its weight during teleportation, hence the notation \emph{recorded link teleportation}.

Introducing (\ref{pref}) into the formal solution (\ref{formalB}) leads to the expression
 \begin{align}
\label{formalK}
\pi_{i;\alpha} = \frac{w_i^{\rm in}}{W} + \sum_{k=1}^{\infty} \frac{\alpha^k}{W}  \sum_j  T^k_{ij} \left( w_j^{\rm in}- w_j^{\rm out}  \right),
\end{align}
which differs from (\ref{formalK2}) in several ways. At zeroth order, PageRank for recorded link teleportation is not uniform anymore, and it is simply given by in-strength, which is itself a standard and widely-used centrality measure. $k^{th}$-order contributions are made of a weighted average of contributions at path of length $k$. The contribution of each node, instead of each link for (\ref{formalK2}), is the difference between its in-strength and its out-strength, $w_j^{\rm in} - w_j^{\rm out}$. As expected, nodes concentrating the flow of probability, $w_j^{\rm in} > w_j^{\rm out}$ give a positive contribution, while nodes diluting this flow, $w_j^{\rm in} < w_j^{\rm out}$, give a negative contribution. Equation (\ref{formalK}) thus interpolates between local and non-local centrality measures when tuning $\alpha$. By construction, (\ref{formalK}) also has the interesting property that the PageRank vanishes for leaves, i.e., nodes whose in-strength is equal to zero, for any $\alpha<1$, in agreement with PageRank's original philosophy that votes come from in-neighbours.

Recorded link teleportation offers a range of interesting mathematical properties that make it an ideal candidate for our purpose.
Contrary to recorded node teleportation (\ref{formalK2}), in which PageRank depends on $\alpha$ except in trivial situations, (\ref{formalK}) has the advantage of being independent of $\alpha$ when the network is undirected or when the network is Eulerian ($w_i^{\rm in}=w_i^{\rm out}$, $\forall i$), as (\ref{formalK}) obviously reduces to $\pi_{i} = w_i^{\rm in}/W$ in those cases. Additionally, it is straightforward to show that PageRank is also given by $\pi_{i}=w_{i}^{\rm in}/W$ for any $\alpha$ in the mean-field approximation, where the adjacency matrix takes the form $W_{ij} \approx w_i^{\rm in} w_j^{\rm out}/W$, as can be checked from Eq.~(\ref{preference})
\begin{align}
\frac{w_i^{\rm in}}{W} = \alpha \sum_{j} \frac{w_i^{\rm in} w_j^{\rm out}}{W w_j^{\rm out}}  \frac{w_j^{\rm in}}{W} + (1-\alpha)  \frac{w_i^{\rm in}}{W}.
\end{align}
This result is expected to hold in large, well-mixed networks, where mean-field approximations are known to provide reasonable predictions. Taken together, these results thus suggest that recorded link teleportation, by blending the directed and the undirected solutions instead of trading off the directed solution with the uniform solution, provides rankings that are more robust to the exact value of the teleportation rates.

\subsubsection*{Unrecorded teleportation}
Despite its apparent success at minimizing the effects of teleportation on the value of PageRank, recorded link teleportation suffers from an important limitation: all transitions are treated equal. This property has unwanted consequences when using random walks to uncover communities in a network, as teleportation tends to create artificial connections between nodes in different communities, and thus to water down structures present in the system. In order to circumvent this limitation, we propose the concept of unrecorded teleportation, where only steps along links are considered when performing a measure of the network \cite{bergstrom2008eigenfactor,rosvall2011multilevel}.

The stationary solution of a random walk with unrecorded teleportation process can easily be calculated by applying an extra step {\em without} teleportation to the solution for the corresponding recorded teleportation
 \begin{align}
\label{formal2}
\pi_{i;\alpha}^{\rm unrec} = \sum_l T_{il} \pi_{l;\alpha},
\end{align}
which leads to the expression
 \begin{align}
\label{formal3}
\pi_{i;\alpha}^{\rm unrec} = (1-\alpha) \sum_l T_{il} \sum_j (I-\alpha T)^{-1}_{lj} v_j,
\end{align}
and to the Taylor expansion
\begin{align}
\label{formal4}
\pi_{i;\alpha}^{\rm unrec} = \sum_l T_{il} \left(v_l + \sum_{k=1}^{\infty} \alpha^k  \sum_j \left( T^k_{lj} - T^{k-1}_{lj}  \right) v_j \right),
\end{align}
the behavior of which depends on the choice of $v_j$. In the following, we consider two versions of unrecorded teleportation. 

In the first version, shown in Fig.~\ref{schematicwalk}(c), called unrecorded node teleportation, the preference vector is uniform. Unfortunately, this version does share the aforementioned robust properties of recorded link teleportation for PageRank. It is nonetheless interesting to note that the leading contribution for small values of $\alpha$ is given by 
 \begin{align}
\pi_{i;\alpha}^{\rm unrec} = \sum_j T_{ij} + O(\alpha),
\end{align}
which simply counts the weight of incoming links normalized by the out-strength of the neighbor. This centrality measure finds potential applications in bibliometrics, as it takes into account the variability in the number of references (out-links) per article, and should facilitate comparisons of scientific journals and authors across scientific fields \cite{moed1985use}.

Unrecorded link teleportation, shown in Fig.~\ref{schematicwalk}(d), is defined by a preference vector proportional to out-strength. This choice has the advantage of effectively leading to a sampling of the links proportionally to their weight in the network. Indeed, selecting a node with a probability proportional to its out-strength and following one of its links before recording is equivalent to selecting a link at random in the network. This equivalence ensures that the stationary solutions of random walks with recorded link teleportation and with unrecorded link teleportation are identical, and are given by Eq.(\ref{formalK}). Unrecorded link teleportation thus presents the same robustness, e.g., independence of PageRank on $\alpha$ for undirected networks, in the mean-field approximation, etc. As we will see in simulations in the next section, unrecorded link teleportation has the further advantage of stabilizing the outcome of community detection algorithms applied to real and artificial benchmark networks.

\subsubsection*{Smart teleportation and clustering}

\begin{figure}[tpb]
    \centering
   \includegraphics[width=\columnwidth]{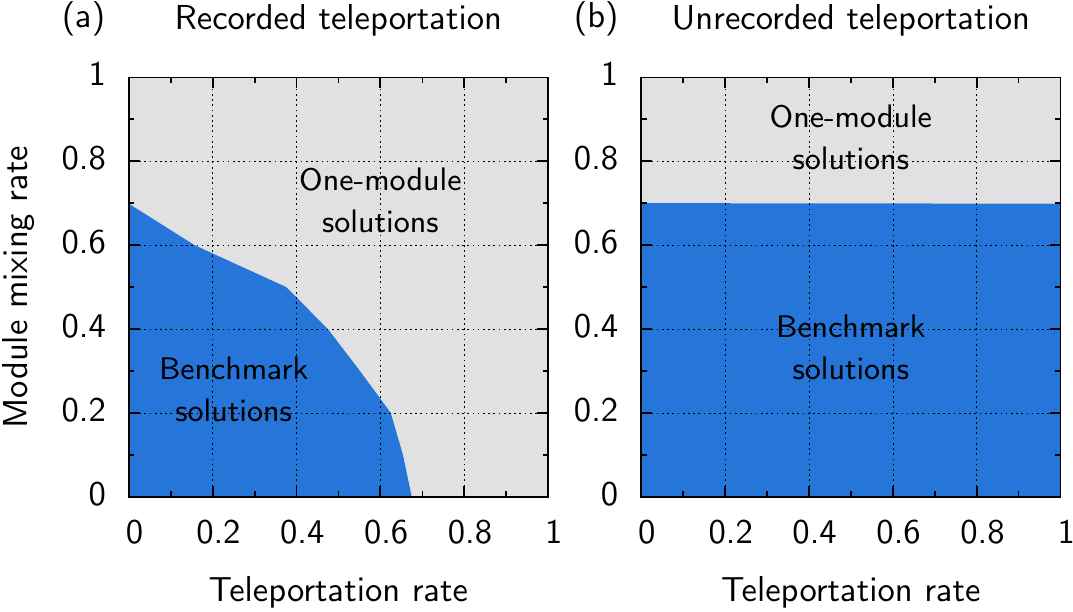}
    \caption{.pdfRobust clustering with unrecorded teleportation in directed benchmark networks. In the blue region, the benchmark solution with multiple modules minimizes the map equation for recorded (a) and unrecorded teleportation (b). In the gray region, the one-module solutions is optimal for recorded (a) and unrecorded teleportation (b). We used Infomap to minimize directed LFR benchmark networks with 1,000 nodes and 7,500 links with between 20 and 50 nodes in the communities \cite{lancichinetti2009community}. Results do not depend on the teleportation target in the benchmark networks without degree-degree correlations and with uniform out-degree.
\label{mapteleportationlimit}}
\end{figure}

To explain why unrecorded teleportation gives much more robust partitions than recorded teleportation, we have partitioned unweighted directed benchmark networks with known partitions and tunable module mixing rates $\mu$ \cite{lancichinetti2009community}. Figure \ref{mapteleportationlimit} shows that Infomap either finds the benchmark solution or leaves the benchmark unpartitioned in one single module for all teleportation schemes. Since the random walker movements between modules are only marginally affected by the target of teleportation in the benchmark networks without degree-degree correlations and with uniform out-degree, results obtained from link teleportation and node teleportation are practically the same. But clustering obtained with unrecorded or recorded teleportation makes all the difference as shown in Fig.~\ref{mapteleportationlimit}. If teleportation steps are not encoded, the results become independent of teleportation rate for a given module mixing and Infomap recovers the benchmark solution for all module mixings up until $\mu=0.7$. At the same mixing rate, with 70 percent of each node's links connecting to nodes outside its own cluster, recorded teleportation hits the limit for which no low teleportation rate can generate the benchmark solution. Above module mixing rate $\mu=0.7$, the community structure of the benchmark network is smeared out by any teleportation rate. For module mixing rates approaching zero, Infomap can recover the benchmark solutions at increasing teleportation rates. 

\begin{figure}[tpb]
    \centering
   \includegraphics[width=\columnwidth]{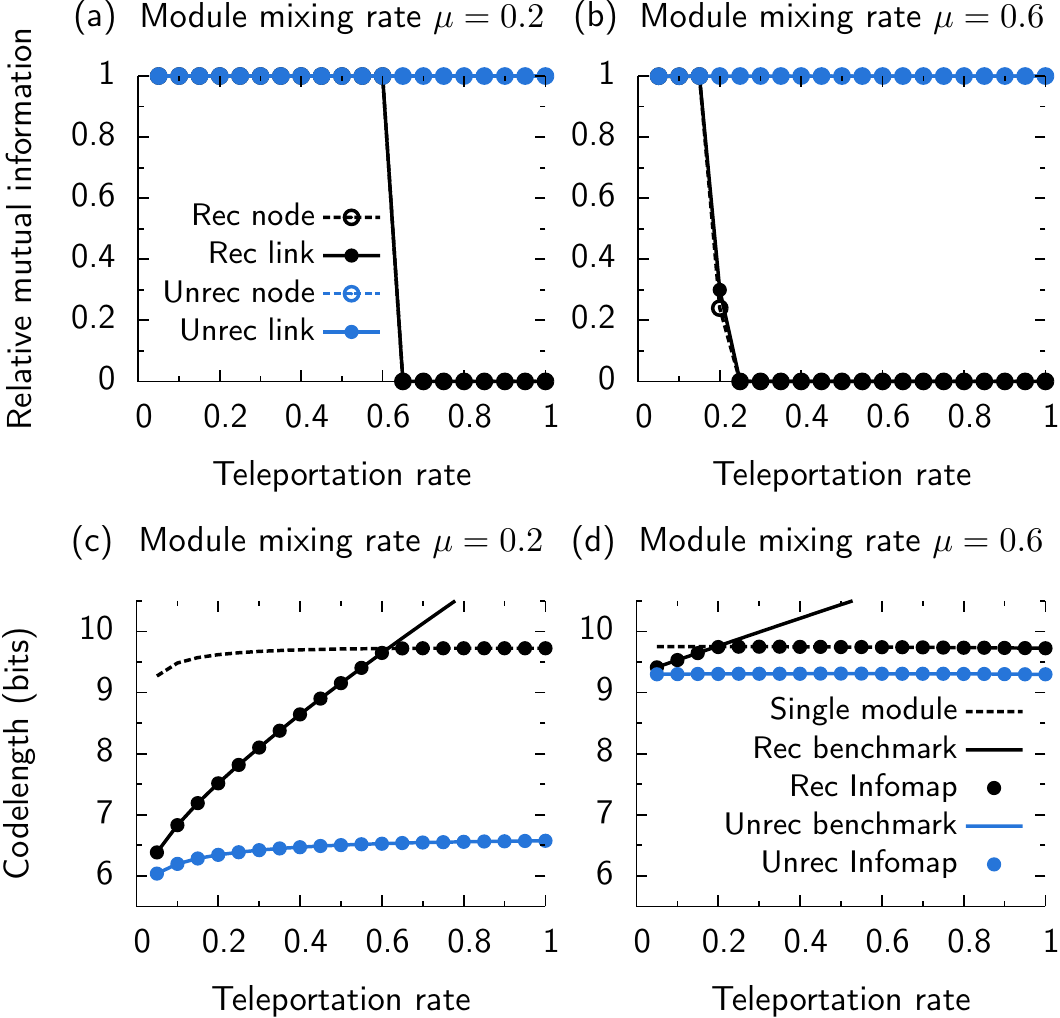}
    \caption{.pdfUnrecorded teleportation strongly reduces the influence of the teleportation rate. The top panes show the normalized mutual information between the benchmark partition at low module mixing rate (a) and high module mixing rate (b), and partitions generated by Infomap for four different teleportation schemes: without encoding of teleportation steps to links (Unrec link) and nodes (Unrec node) and with encoding of teleportation steps to links (Rec link) and nodes (Rec node). The bottom panes show the codelength of the map equation for different partitions of benchmark networks at low module mixing rate (c) and high module mixing rate (d): the codelengths associated with unpartitioned networks (with and without encoding of teleportation steps) and the codelengths associated with the benchmark partition and the partition generated by Infomap (with and without encoding of teleportation steps). Results are based on same data as in Fig.~\ref{mapteleportationlimit}. \label{mapteleportation}}
\end{figure}

Figure \ref{mapteleportation} explains what partition Infomap will find at different module mixing and teleportation rates. As the figure shows, Infomap finds the benchmark partition as long as the benchmark partition provides a shorter description length than the unpartitioned network. The sharp transition from the benchmark solution with multiple modules to the unpartitioned one-module solution happens without any intermediate solutions at teleportation rate $1-\alpha = 0.6$ for module mixing rate $\mu = 0.2$, as shown in Figs.~\ref{mapteleportation}(a) and (c), and at teleportation rate $1-\alpha = 0.2$ for module mixing rate $\mu = 0.6$, as shown in Figs.~\ref{mapteleportation}(b) and (d). Consequently, for recorded, teleportation the total module mixing from links and teleportation determines which solution provides the shortest description of the random walker on the network. For unrecorded teleportation, however, the codelength becomes almost independent of teleportation rate and the module mixing from links determines the clustering result.

In the next section, we will demonstrate the advantages of using smart teleportation over standard teleportation for ranking and clustering real-world networks. 
Contrary to the benchmark networks analyzed above, real-world networks have modules with varying degree of mixing. Therefore, we will not see the sharp transition from a single multi-module solution to the unpartitioned one-module solution. Instead, we will see a gradually decreasing normalized mutual information as the increased teleportation rate smears out the boundaries of weak clusters.

\section*{\large Smart teleportation in real-world networks}

\subsubsection*{Ranking of scientific journals}

\begin{figure}[hpt]
    \centering
   \includegraphics[width=\columnwidth]{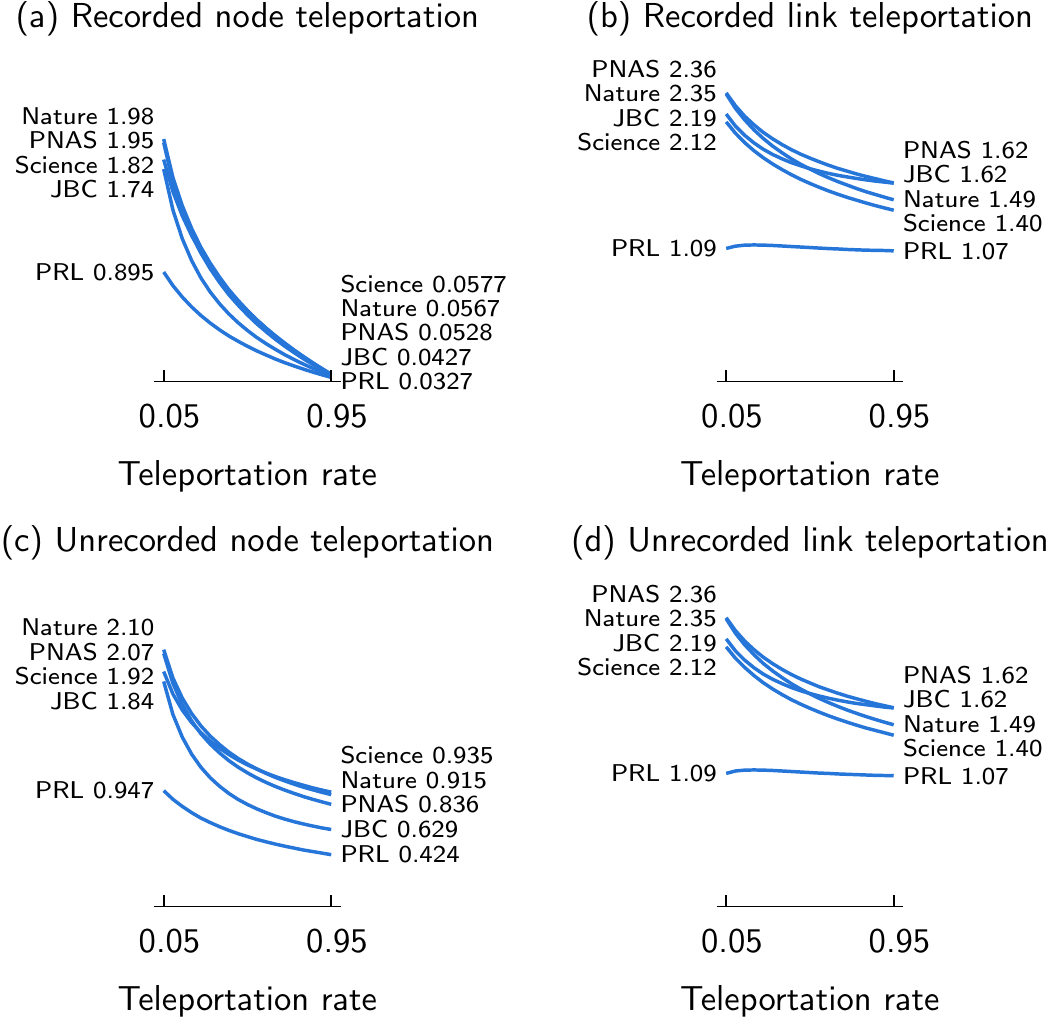}
    \caption{.pdfLink teleportation reduces the influence of teleportation rate on top-ranked scientific journals. Reported in percent, the journal visit rates obtained with recorded node teleportation in (a), recorded link teleportation in (b), unrecorded node teleportation in (c), and unrecorded link teleportation in (d).\label{topranked}}
\end{figure}

In this section we explore the effect of teleportation on ranking and clustering in real-world networks. We begin with an illustrative example of ranking. We ranked 7,940 journals connected by 9.2 million citations aggregated in 1.2 million weighted links \cite{jsrnote} with the four different teleportation schemes for different teleportation rates $1-\alpha$ and reported the node visit rates for five top journals: \emph{Nature}, \emph{Science}, \emph{Proceedings of the National Academy of Sciences} (PNAS), \emph{The Journal of Biological Chemistry} (JBC), and \emph{Physical Review Letters} (PRL) (see Figure \ref{topranked}). The choice of teleportation scheme affects not only the absolute node visit rates, but also the relative node visit rates between the journals, and therefore also the rank order. Teleportation to links, whether recorded or unrecorded, dramatically reduces its undesirable damping effect. Ranking with link teleportation is less sensitive to the choice of teleportation rate, but the rank order nevertheless depends on whether only the local neighborhood of a node for high teleportation rates or the entire network for low teleportation rates is considered. For example, Figs.~\ref{topranked}(b) and \ref{topranked}(d) show that \emph{Nature} ranks higher than \emph{The Journal of Biological Chemistry} only when the entire network structure is taken into account.

Similarly, we find that the choice of teleportation scheme dramatically affects the clustering results. For example, when we clustered the scientific journals with the Infomap method \cite{rosvall2008maps}, we obtained non-trivial solutions for teleportation rates below 50 percent for recorded node teleportation and below 75 percent for recorded link teleportation. For unrecorded teleportation, however, we obtained non-trivial solutions for all teleportation rates.

\subsubsection*{Data description}

To quantitively compare the ranking and clustering results between standard and smart ranking in a more systematic way, we analyzed eight real-world networks. We selected networks of widely different sizes, topologies, and origins:
\begin{list}{}{\leftmargin=1em}
\item[] \emph{Coathorship} is a weighted undirected network included for reference that describes more than 2,500 coauthorships between about 500 network scientists \cite{rosvall2011multilevel}.
\item[] \emph{US airports} is a weighted directed network that describes about 18,000 connections weighted by passenger flow between close to 500 airports in the US in 2007 \cite{rosvall2011multilevel}.
\item[] \emph{US political blogs} is an unweighted directed network that describes about 19,000 hyperlinks between almost 1,500 blogs blogs on US politics collected in 2005 \cite{adamic2005political}. 
\item[] \emph{Swe political blogs} is a weighted directed network that describes about 13,000 connections between more than 1,000 political blogs in Sweden in 2010 \cite{rosvall2011multilevel}. 
\item[] \emph{Journal citations} is a weighted directed network that describes more than a million connections formed by around 10 million citations between close to 8,000 scientific journals in 2007  \cite{rosvall2010mapping}. 
\item[] \emph{Call graph} is a weighted directed network that describes more than 7,000 calls between about 2,500 functions in  the cross-platform library GLib \cite{rosvall2011multilevel}. 
\item[] \emph{Stanford web} is a directed network that describes 2.3 million hyperlinks between almost 300,000 web pages in the domain stanford.edu \cite{leskovec2009community}.
\item[] \emph{Google web} is a directed network that describes 5.1 million hyperlinks between more than 700,000 web pages from the Google Programming Contest in 2002 \cite{leskovec2009community}. 
\end{list}

\begin{figure*}[t]
\centering
\includegraphics[width=0.8\textwidth]{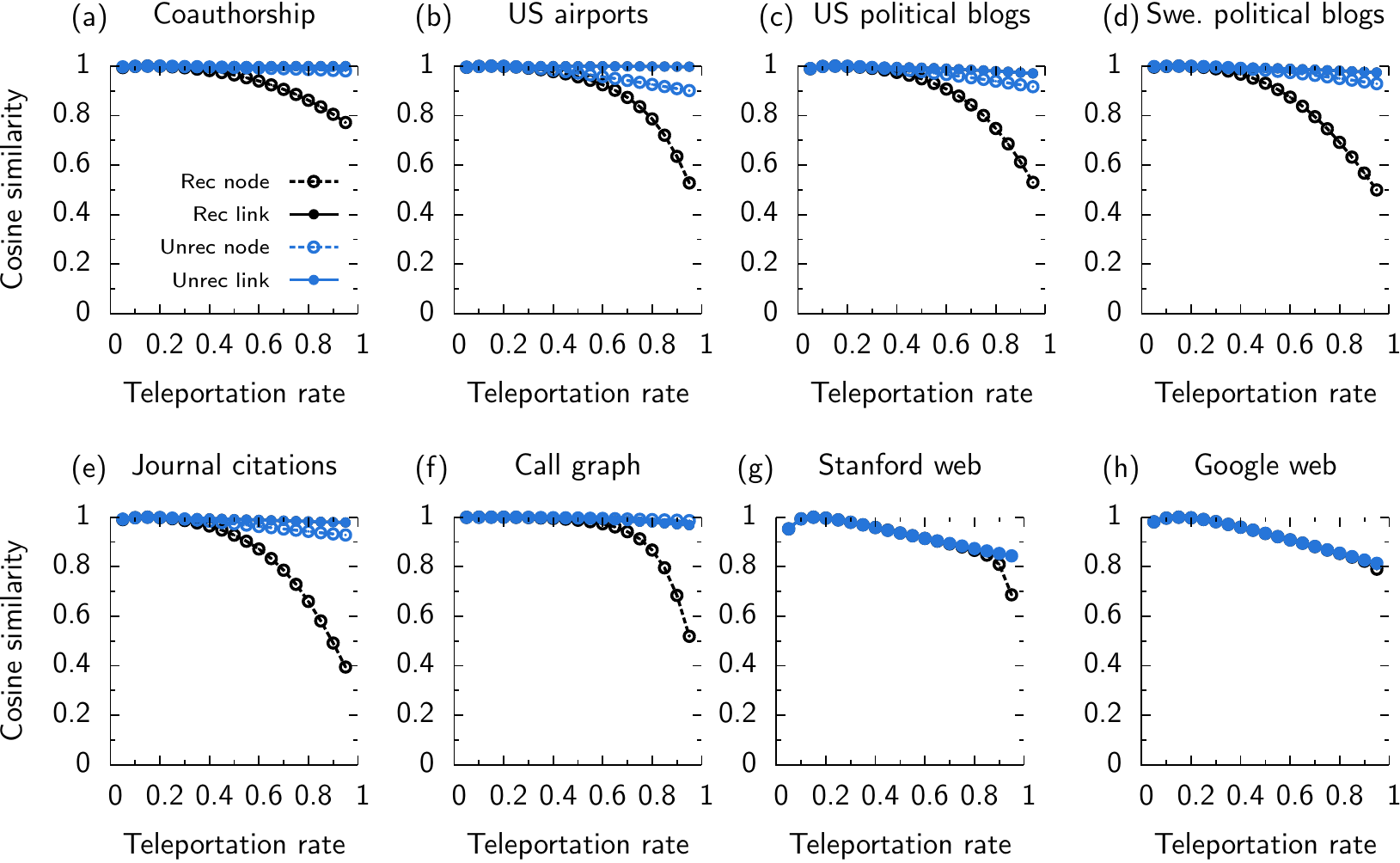}
\caption{.pdfRobust rank size with smart teleportation in real-world networks. We measured the cosine similarity between the node rank sizes obtained at teleportation rate $1-\alpha=0.15$ and the node rank sizes obtained at lower and higher teleportation rates. When teleportation steps were included in the node visit rates of the random walker, teleportation to links (Rec link) is more robust than uniform teleportation to nodes (Rec node). When teleportation steps were not included in the node visit rates, the rank size is overall more robust and link teleportation (Unrec link) is again more robust than node teleportation (Unrec node). Note that recorded and unrecorded link teleportation by definition give the same rank size. See main text for details about the networks.\label{realsizeteleportation}}
\end{figure*}

\begin{figure*}[t]
\centering
\includegraphics[width=0.8\textwidth]{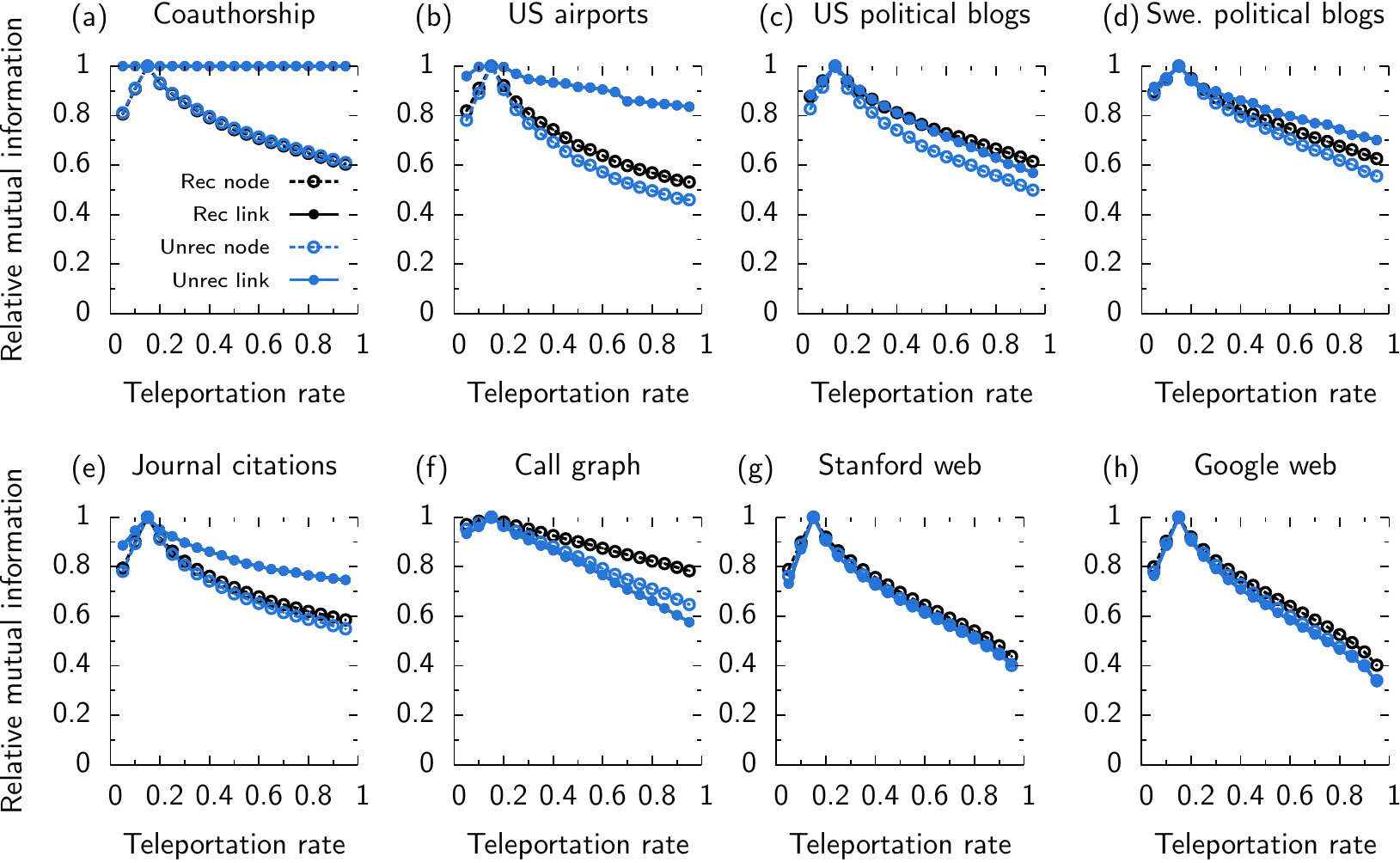}
\caption{.pdfRobust rank order with smart teleportation in real-world networks. We measured the mutual information between the node rank order obtained at teleportation rate $1-\alpha=0.15$ and the node rank order obtained at lower and higher teleportation rates. In general, there is no advantage in not counting teleportation steps (Unrec) over counting teleportation steps (Rec), but link teleportation (link) is again more robust than node teleportation (node). Note that recorded and unrecorded link teleportation by definition give the same rank order. See main text for details about the networks.\label{realrankteleportation}}
\end{figure*}

For each network, we analyzed ranking and clustering robustness of the four different teleportation schemes depicted in Fig.~\ref{schematicwalk}: recorded teleportation to nodes and to links and unrecorded teleportation to nodes and to links. We quantified the robustness of the results to variations in the teleportation rate by measuring the similarity between results obtained at the commonly used teleportation rate $1-\alpha=0.15$ with results obtained at lower and higher teleportation rates. 

\subsection*{Robust ranking}

We used the power iteration method to derive the node visit frequencies for the four different teleportation schemes. When located on a node with $k^{\rm out}_i =0$, a random walker automatically performs a teleportation, as in the original formulation PageRank. 
To obtain the node visit frequencies for the unrecorded teleportation scheme, we first calculated the node visit frequencies with the recorded teleportation scheme and then performed an extra step without teleportation followed by normalization. We are interested in the robustness of both the node rank sizes and the node rank order. There are several ways to measure the similarity between the sizes and orderings of two node rankings, but we opted for two simple metrics.

For rank size comparisons between different node visit rates $\pi_{i;x}$ and $\pi_{i;y}$ obtained by different teleportation rates $1-\alpha_x$ and $1-\alpha_y$, we used the commonly used cosine similarity
\begin{align}
	S = \frac{\sum_{i}\pi_{i;x}\pi_{i;y}}{\sqrt{\sum_{i}\pi_{i;x}^{2}}\sqrt{\sum_{i}\pi_{i;y}^{2}}}.
\end{align}
Figure \ref{realsizeteleportation} shows the cosine similarity measured between the node rank sizes obtained at teleportation rate $1-\alpha=0.15$ and the node rank sizes obtained at lower and higher teleportation rates. As for the top-ranked journals in Fig.~\ref{topranked}, for all teleportation schemes the results depend on the teleportation rate and the similarity is only perfect at the reference teleportation rate $1-\alpha=0.15$. But for all networks, link teleportation is equally or more robust than node teleportation, and unrecorded teleportation is equally or more robust than recorded teleportation. The commonly used recorded teleportation to nodes is by far the least robust teleportation scheme.

\begin{figure*}[t]
    \centering
   \includegraphics[width=0.8\textwidth]{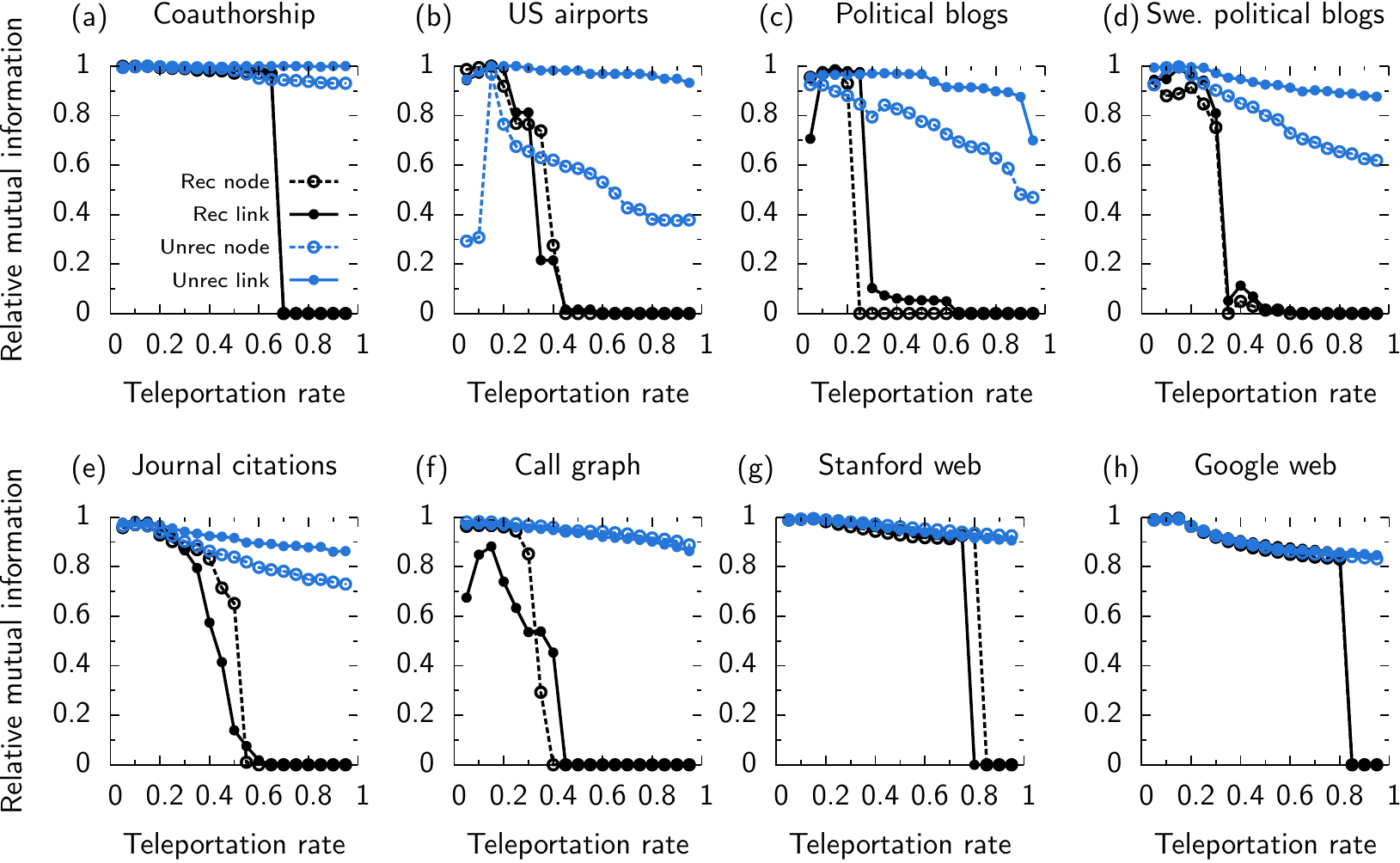}
    \caption{.pdfRobust clustering without encoding of teleportation steps in real-world networks. We measured the mutual information between the obtained partitions at teleportation rate $1-\alpha=0.15$ and the obtained partitions at lower and higher teleportation rates. Not encoding teleportation steps (Unrec) is always better than encoding teleportation steps (Rec). The robustness of teleportation to nodes (node) or links (link) depends on the network. Each data point corresponds to the average over 100 pairwise comparisons between partitions generated with the Infomap method \cite{rosvall2008maps}. See main text for details about the networks. \label{realmapteleportation}}
\end{figure*}

For comparing different node rank orders, we measured the mutual information between node-pair comparisons sampled from the rankings. That is, we sampled pairs $i,j$ of nodes proportional to the node rank sizes $\pi_{i;x},\pi_{j;x}$ from the ranking obtained at teleportation rate $1-\alpha_x$ and measured the reduction of uncertainty about which of the two nodes $X\text{=}\{i,j\}$ has the highest rank after observing the order $Y$ of the other ranking obtained at teleportation rate $1-\alpha_y$. In general, with joint probability distribution $p(x,y)$ and marginal probability distributions $p(x) = \sum_y p(x,y)$ and $p(y) = \sum_x p(x,y)$, the mutual information is given by 
\begin{align}
	I = \sum_{x,y}p(x,y)\log\left(\frac{p(x,y)}{p(x)p(y)}\right).
\end{align}
With the unit step function
\begin{align}
	\theta(z) = 
	\begin{cases}
		1 & \text{if } z \ge 0 \\
		0 & \text{if } z < 0
	\end{cases},
\end{align}
the joint probability of, for example, $X=i$ and $Y=j$, is 
\begin{align}
p(i,j) = \sum_{i,j} \theta(\pi_{i;x}-\pi_{j;x}) \theta(\pi_{j;y}-\pi_{i;y}) \pi_{i;x}\pi_{j;x}.
\end{align}
The factor $\pi_{i;x}\pi_{j;x}$, obtained by picking nodes proportional to their visit frequencies, guarantees that the order between highly ranked nodes weighs higher in the comparison.  
If one ranking provides no information about the other ranking, one bit of information would be necessary to determine which of two nodes is the one with the highest rank. Therefore, the mutual information can not be larger than one bit. But because some pair of nodes in general can have the same rank, we normalize the mutual information by dividing by the maximum entropy of $X$ and $Y$. With the entropy given by
\begin{align}
-\sum_x p(x) \log p(x),
\end{align}
the normalized mutual information takes the form
\begin{align}
	R = \frac{I(X;Y)}{\max\left(H(X),H(Y)\right)}. \label{EqMI}
\end{align}
We normalize by dividing by the maximum entropy of $X$ and $Y$ rather than the commonly used average to avoid rewarding simplistic solutions with many or all nodes of equal rank.
Figure \ref{realrankteleportation} shows the normalized mutual information between the node rank order obtained at teleportation rate $1-\alpha=0.15$ and the node rank obtained at lower and higher teleportation rates. 
For all networks, rank order is the same for unrecorded and recorded teleportation when teleporting to links as shown in Fig.~\ref{schematicwalk}.
The rank order generated by node teleportation is more robust in the strongly directed call graph, but more often the rank order generated by link teleportation is more robust. For example, all link teleportation rates generate the same rank order in the undirected coauthorship network, whereas the rank order is influenced by node teleportation rates. Teleportation to links can take advantage of possible bidirectional connections between nodes.

\subsection*{Robust clustering}

When clustering nodes in networks based on random walks, not only the node visit rates, as for ranking, but also the transition rates between nodes affect the result. Therefore, and as the example with scientific journals above demonstrates, the teleportation scheme and teleportation rate have dramatic effects on clustering. To quantitatively compare the teleportation schemes, we clustered the eight real-world networks with the information-theoretic clustering method Infomap \cite{rosvall2008maps}. Infomap searches for the network partition that minimizes the description length of a random walker guided by the links of the network. By altering the dynamics of the random walker, for example, by altering the teleportation rate, Infomap may consequently identify different partitions. To test how much the partitions change when  the teleportation rate is altered, we used the normalized mutual information applied to cluster comparisons \cite{danon2005comparing}. In this way, we can compare the robustness associated with the different teleportation schemes.

The mutual information between two network partitions measures how much we learn about one network partition by studying the other one. We always used the network partition obtained at the commonly used teleportation rate $1-\alpha = 0.15$ as reference. To avoid undesirable effects that singletons can cause, we sampled the nodes proportionally to their visit frequencies rather than uniformly when calculating mutual information. In this way, we also put emphasis on correct assignments of important nodes. For fair comparisons between partitions obtained at different teleportation rates, we used the node visit frequencies of the reference partition. By normalizing by the maximum entropy of the two partitions rather than the average of the two, we naturally penalize for overfitting and avoid rewarding for underfitting.

With $s_x$ for the total visit rate of all nodes in module $x$ and $s_{xy}$ for the total visit rate of all nodes that are jointly partitioned in module $x$ and module $y$, the entropy takes the form
\begin{align}
	H(X) = -\sum_{x} s_x\log s_x,
\end{align}
the mutual information
\begin{align}
I(X;Y) = \sum_{x,y} s_{xy}\log \frac{s_{xy}}{s_x s_y},
\end{align}
and the normalized mutual information as in Eq.\ (\ref{EqMI}).

Figure \ref{realmapteleportation} shows that unrecorded teleportation gives more robust clustering for all tested networks and, in general, link teleportation gives more robust results than node teleportation. Recorded teleportation gives robust results in a window around teleportation rate $1-\alpha=0.15$, but the normalized mutual information quickly drops to zero outside this window. Contrarily, for unrecorded teleportation, the normalized mutual information stays relatively high for all values of the teleportation rate.

\section*{\large Conclusions}

When ranking and clustering nodes in networks, we have demonstrated analytically and numerically that we can drastically reduce undesirable and parameter-dependent effects of standard teleportation with unrecorded teleportation to links. Because this smart teleportation scheme takes advantage of the topology of the network --- blending the directed and the undirected solutions instead of trading off the directed solution with the uniform solution --- results are more robust to the exact value of the  teleportation rates. In particular, we have shown analytically that ranking results are exact and independent of teleportation rates for undirected and well-mixed networks, and for all the real-world networks we have analyzed, smart teleportation is as good as or better than standard teleportation.

When clustering networks with Infomap based on the movements of a random walker on the network, not recording the teleportation steps makes the results of real-world networks dramatically more robust. Because smart teleportation eliminates mixing between network communities, results of benchmark networks are practically independent of the teleportation rate. The advantages of smart teleportation over standard teleportation makes it interesting to explore the benefits in other flow-based clustering algorithms and variations of PageRank.

\begin{acknowledgments}
We are grateful to Jevin West for providing many helpful comments and suggestions. MR was supported by the Swedish Research Council grant 2009-5344.
\end{acknowledgments}	


\begin{thebibliography}{10}

\bibitem{boccaletti2006complex}
S.~Boccaletti, V.~Latora, Y.~Moreno, M.~Chavez, and D.~Hwang,
\newblock Physics reports {\bf 424}, 175 (2006).

\bibitem{brin1998anatomy}
S.~Brin and L.~Page,
\newblock Computer networks and ISDN systems {\bf 30}, 107 (1998).

\bibitem{langville2006google}
A.~Langville and C.~Meyer,
\newblock {\em Google page rank and beyond} (Princeton Univ Pr, 2006).

\bibitem{haveliwala2003topic}
T.~Haveliwala,
\newblock Knowledge and Data Engineering, IEEE Transactions on {\bf 15}, 784
  (2003).

\bibitem{Goncalves09wsdm}
B.~Goncalves, M.~Meiss, J.~Ramasco, A.~Flammini, and F.~Menczer,
\newblock Remembering what we like: Toward an agent-based model of web traffic,
\newblock in {\em Late-breaking result at WSDM}, 2009.

\bibitem{delvenne2011centrality}
J.~Delvenne and A.~Libert,
\newblock Physical Review E {\bf 83}, 046117 (2011).

\bibitem{bergstrom2008eigenfactor}
C.~Bergstrom, J.~West, and M.~Wiseman,
\newblock The Journal of Neuroscience {\bf 28}, 11433 (2008).

\bibitem{chen2007finding}
P.~Chen, H.~Xie, S.~Maslov, and S.~Redner,
\newblock Journal of Informetrics {\bf 1}, 8 (2007).

\bibitem{allesina2009googling}
S.~Allesina and M.~Pascual,
\newblock PLoS computational biology {\bf 5}, e1000494 (2009).

\bibitem{radicchi2011best}
F.~Radicchi,
\newblock PloS one {\bf 6}, e17249 (2011).

\bibitem{rosvall2008maps}
M.~Rosvall and C.~Bergstrom,
\newblock Proceedings of the National Academy of Sciences {\bf 105}, 1118
  (2008).

\bibitem{delvenne2010stability}
J.~Delvenne, S.~Yaliraki, and M.~Barahona,
\newblock Proceedings of the National Academy of Sciences {\bf 107}, 12755
  (2010).

\bibitem{borgatti2005centrality}
S.~Borgatti,
\newblock Social Networks {\bf 27}, 55 (2005).

\bibitem{lambiotte2011flow}
R.~Lambiotte {\em et~al.},
\newblock Physical Review E {\bf 84}, 017102 (2011).

\bibitem{boldi2005pagerank}
P.~Boldi, M.~Santini, and S.~Vigna,
\newblock Pagerank as a function of the damping factor,
\newblock in {\em Proceedings of the 14th international conference on World
  Wide Web}, pp. 557--566, ACM, 2005.

\bibitem{brinkmeier2006pagerank}
M.~Brinkmeier,
\newblock ACM Transactions on Internet Technology (TOIT) {\bf 6}, 282 (2006).

\bibitem{pretto2002theoretical}
L.~Pretto,
\newblock A theoretical analysis of google's pagerank,
\newblock in {\em String Processing and Information Retrieval}, pp. 125--136,
  Springer, 2002.

\bibitem{langville2004deeper}
A.~Langville and C.~Meyer,
\newblock Internet Mathematics {\bf 1}, 335 (2004).

\bibitem{avrachenkov2008singular}
K.~Avrachenkov, N.~Litvak, and K.~Pham,
\newblock Internet Mathematics {\bf 5}, 47 (2008).

\bibitem{baeza2006generalizing}
R.~Baeza-Yates, P.~Boldi, and C.~Castillo,
\newblock Generalizing pagerank: damping functions for link-based ranking
  algorithms,
\newblock in {\em Proceedings of the 29th annual international ACM SIGIR
  conference on Research and development in information retrieval}, pp.
  308--315, ACM, 2006.

\bibitem{fortunato2008approximating}
S.~Fortunato, M.~Bogu{\~n}{\'a}, A.~Flammini, and F.~Menczer,
\newblock Algorithms and Models for the Web-Graph , 59 (2008).

\bibitem{litvak2007degree}
N.~Litvak, W.~Scheinhardt, and Y.~Volkovich,
\newblock Internet mathematics {\bf 4}, 175 (2007).

\bibitem{rosvall2011multilevel}
M.~Rosvall and C.~Bergstrom,
\newblock PloS One {\bf 6}, e18209 (2011).

\bibitem{moed1985use}
H.~Moed, W.~Burger, J.~Frankfort, and A.~Van~Raan,
\newblock Research Policy {\bf 14}, 131 (1985).

\bibitem{lancichinetti2009community}
A.~Lancichinetti and S.~Fortunato,
\newblock Physical Review E {\bf 80}, 056117 (2009).

\bibitem{jsrnote}
Thomson-Reuters’ Journal Citation Reports 2007. Our data tally on a
  journal-by-journal basis the citations from articles published in 2007 to
  articles published in the previous two years. Because we are interested in
  relationships between journals, we exclude journal self-citations.

\bibitem{adamic2005political}
L.~Adamic and N.~Glance,
\newblock The political blogosphere and the 2004 us election: divided they
  blog,
\newblock in {\em Proceedings of the 3rd international workshop on Link
  discovery}, pp. 36--43, ACM, 2005.

\bibitem{rosvall2010mapping}
M.~Rosvall and C.~Bergstrom,
\newblock PloS one {\bf 5}, e8694 (2010).

\bibitem{leskovec2009community}
J.~Leskovec, K.~Lang, A.~Dasgupta, and M.~Mahoney,
\newblock Internet Mathematics {\bf 6}, 29 (2009).

\bibitem{danon2005comparing}
L.~Danon, A.~D{\'\i}az-Guilera, J.~Duch, and A.~Arenas,
\newblock Journal of Statistical Mechanics: Theory and Experiment {\bf 2005},
  P09008 (2005).

\end{thebibliography}

\end{document}